# Using off-treatment sequential multiple imputation for binary outcomes to address intercurrent events handled by a treatment policy strategy

Sunita Rehal[1], Nicky Best[1], Sarah Watts[1], Thomas Drury[1]

[1]GlaxoSmithKline, London, UK

## Abstract

The estimand framework proposes different strategies to address intercurrent events. The treatment policy strategy seems to be the most favoured as it is closely aligned with the pre-addendum intention-to-treat principle. All data for all patients should ideally be collected, however, in reality patients may withdraw from a study leading to missing data. This needs to be dealt with as part of the estimation. Several areas of research have been conducted exploring models to estimate the estimand when intercurrent events are handled using a treatment policy strategy, however the research is limited for binary endpoints.

We explore different retrieved dropout models, where post-intercurrent event, the observed data can be used to multiply impute the missing post-intercurrent event data. We compare our proposed models to a simple imputation model that makes no distinction between the pre- and post-intercurrent event data, and assess varying statistical properties through a simulation study. We then provide an example how retrieved dropout models were used in practice for Phase 3 clinical trials in rheumatoid arthritis.

From the models explored, we conclude that a simple retrieved dropout model including an indicator for whether or not the intercurrent event occurred is the most pragmatic choice. However, at least 50% of observed post-intercurrent event data is required for these models to work well. Therefore, the suitability of implementing this model in practice will depend on the amount of observed post-intercurrent event data available and missing data.

# Introduction

The ICH E9(R1) addendum(1) introduces a framework that aims to translate the trial objective to a precisely defined and clearly understood treatment effect which is labelled "an estimand". Only once the clinical objective and estimand are determined, should a suitably aligned estimator be chosen.

Intercurrent events (IEs) are defined in the addendum as "events occurring after treatment initiation that affect either the interpretation or the existence of the measurements associated with the clinical question of interest". To address the occurrence of IEs, the addendum proposes five different strategies that can be used which provide context on how the IEs are reflected in the estimand defined. We focus on the treatment policy strategy which is used to estimate the effect of assignment to treatment. This strategy is often considered to better reflect what may occur in clinical practice(2). When this strategy is used, any additional effects of the IE(s) are considered relevant and included in the targeted effect. Importantly, when estimating these types of effects, the outcome measures after the occurrence of an IE need to continue to be collected and used for estimation. Ideally, this should be collected for all patients, but in practice this is not always achievable as patients may withdraw from the study. This leads to missing data after the occurrence of the intercurrent event and complicates the estimation (Figure 1, Profile 3).

Figure 1: Patient profiles when a treatment policy strategy is used.

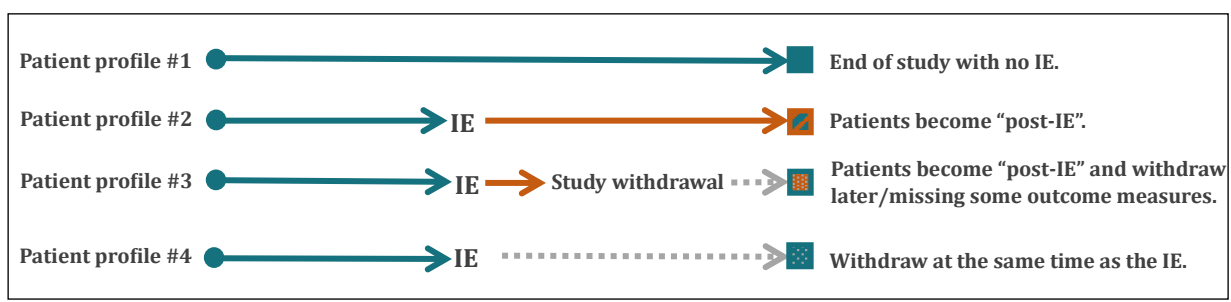

Using multiple imputation (MI) is one approach that could be used for estimation to account for the missing data post-IE. Since the guideline was finalised in 2019, several areas of research using MI have been conducted to estimate the estimand for IEs handled by a treatment policy strategy for recurrent event(3), time-to-event(4) and continuous(2,5–7) endpoints. Irrespective of the type of endpoint and disease area, the results from this research highlight it is important to consider: 1) how much missing data there is and 2) the amount of post-IE data that is available so the targeted effect can be reliably estimated. The results from investigating the MI models and MMRM models for continuous outcomes(6,7) show that there is no single optimal approach, and the choice of which estimation method to use leads to a bias-variance trade off. To date, there has been limited research into estimation methods to handle missing data for estimands with binary outcomes that use the treatment policy strategy to address IEs. While the recent methods on using reference-based MI for binary outcomes has been shown to perform well using a multivariate normal model, the performance of these models are based on considering the data post-IE equivalent to missing data(8).

Where the underlying distribution of a binary outcome is continuous, the general recommendation is to impute based on that underlying continuous distribution(9,10). However, this can be problematic where the outcome is either truly binary or determined from several individual components at several follow up visits leading to nonconvergence of the imputation model(10). A common issue that requires additional consideration to impute binary outcomes is perfect prediction(11). This is prevalent where data are sparse or if a single imputation model is used across multiple visits.

We propose imputation models similar to those used by Roger et al.(3), Hartley et al.(4) and Drury et al.(6) and extend these to impute missing binary data using observed post-IE data. The performance of the proposed models are assessed through two simulation studies. The first investigates the models by varying the proportion of observed post-IE data and missing data, and the second considers different sample sizes under a single scenario. Two clinical trials are then discussed which implemented these types of models in practice.

The remainder of this paper is structured as follows. First, we outline our motivating example for rheumatoid arthritis, we then perform a simulation study under different scenarios to investigate different MI models to target the estimand in the presence of missing data. Finally we show how these types of models can be applied for two example case studies.

# Motivation: The contRAst 1 and contRAst 2 studies

This work is motivated by studies in rheumatoid arthritis (RA). ACR20 response is commonly used as the primary endpoint in this disease. ACR20 measures whether there is a 20% improvement in both tender and swollen joint counts and 20% improvement in three of the five remaining core set measures: patient and physician global, pain, disability, and an acute phase reactant. ACR20 uses several components measured on a continuous scale to calculate a binary response of achieving ACR20 (success) or not (failure). This is a validated endpoint that is accepted by the FDA(12).

ContRAst 1 and contRAst 2(13) were two Phase 3 studies investigating Otilimab at 90mg and 150mg in patients with RA including patients who have inadequate responses to methotrexate (contRAst 1) or conventional synthetic/biologic disease-modifying antirheumatic drugs (cs/bDMARDs; contRAst 2). The studies had also included an active control arm (Tofacitinib). **Trial registration numbers** NCT03980483, NCT03970837.

Both study's primary objective was: To compare the efficacy of Otilimab at doses of 90 mg and 150 mg weekly versus placebo for the treatment of participants with moderately to severely active RA who are on a stable background of csDMARD(s) and who have had an inadequate response to methotrexate (contRAst 1) or cs/bMARDs (contRAst 2). This leads to an associated primary estimand for contRAst 1 that can be described as follows:

*The conditional odds ratio of ACR20 response at Week 12 in Otilimab compared with placebo in patients with moderately to severely active RA despite MTX treatment, including any subsequent effects of IP discontinuation, prohibited medication use or change in stable dose of background therapy.*

The primary estimand for contRAst 2 is similar to the above with a change in the population attribute to: patients with moderately to severely active RA who are on a stable background of csDMARDs and who have had an inadequate response to cs/bDMARDs.

Both studies planned to impute missing post-IE data based on using a pre-specified step-down approach depending on the amount of post-IE data available. The estimation strategy planned to use MI with on-treatment and off-treatment outcome measures to handle the missing data for the main IE addressed by a treatment policy strategy: IP discontinuation.

Inspired by the models used in these studies, and other more complex retrieved dropout MI models (3,4,6), we set out to assess the performance of these models comparing two treatment arms, active and control (see the models section for more detail).

### Estimand

In this simulation study, we consider a situation where there is a single intercurrent event of interest: treatment discontinuation which is handled using a treatment policy strategy for a rheumatoid arthritis study. We provide a generic description of the estimand (e.g. timing of the endpoint):

| **Primary Estimand** | |
|---|---|
| Population | Adult patients with moderate to severe active rheumatoid arthritis. |

| Endpoint/[Variable] | ACR20 response at the final timepoint. |
|---|---|
| Treatment conditions | Active compared to Control. |
| Summary Measure | Conditional odds ratio. |
| Intercurrent Events (strategy) | Treatment discontinuation (treatment policy). |
| **Description** | |
| The conditional odds ratio of ACR20 response at the final timepoint between patients assigned to Active compared to Control in adult patients with moderate to severe active rheumatoid arthritis including any additional effects of discontinuation from assigned treatment. | |
| **Clinical Rationale** | |
| Targeting the effect even if some patients discontinue treatment appropriately reflects prescribed treatment in clinical practice. | |

## Estimation

The standard procedure for MI is first to use the observed (completed) data to estimate the parameters required for the imputation model. Here, a logistic regression model is used to reflect that ACR20 response is a binary outcome. Next, the missing values are predicted based on plausible estimates that are drawn from the joint posterior predictive distribution of the missing data conditional on the observed data, creating a single imputed dataset. To account for the uncertainty of the imputed values, the imputation procedure is repeated several times resulting in M imputed datasets. A standard assumption when using MI is that outcomes follow a basic missing at random (MAR) mechanism. Finally, the analysis model - commonly referred to as the substantive model - is fitted to each (M) imputed dataset and combined using Rubin's rules(14).

As with any type of statistical analysis, sensitivity analysis may be used to test for departures from any modelling assumptions made. It is common to investigate departures from the basic MAR assumption via a tipping point analysis by varying the effect in the missing outcomes. Due to the many components required to determine ACR20 response, imputing based on the ACR20 response alone rather than the individual components is preferred. This approach helps to avoid confusion which component(s) are driving any changes that are seen at the tipping point. To make more direct comparisons with this type of sensitivity analysis, we do not consider imputing each individual component assuming a continuous distribution to create the ACR20 scores, as has been discussed elsewhere(9,10,15,16). Therefore, we only consider ACR20 response at all post-baseline visits for this simulation study.

## Methods

### Notation

We consider a parallel two-arm randomised clinical trial where Z = z denotes the treatment patients are assigned to. The two arms are defined as $z \in \{A, C\}$, meaning patients are assigned to receive Active arm $A$ or Control arm $C$ respectively.

We define $Y_{zij}$ as a binary response (i.e. 1 or 0) for each patient $i$ assigned to treatment z at the current visit $j$ for $j = 0, ..., J$ where $j = 0$ represents baseline and $J$ represents the time of assessment. For the simulation studies explored here, ($J = 3$).

We define $D_i$ as an individual's IE with associated integer visit time $D_i(t) = 1,2,3$. We also create $D_{zij}$ as indicators for the occurrence of the IE for the $i^{th}$ patient by the $j^{th}$ visit when assigned to treatment z. Patients at the current visit that have not had the IE are indicated as being on-treatment ($D_{zij} = 0$) and off-treatment ($D_{zij} = 1$) otherwise.

$P_{zij}$ represents the IE patterns for the $i^{th}$ patient assigned to treatment z up to visit $j$. The patterns are created depending on when the IE occurred. Four patterns are created in this scenario, including a pattern for patients that remain in the study for the whole study duration.

## Models

Following Drury et al(6), the estimation approaches we explore use MI with various MAR assumptions to sequentially impute observations at each timepoint separately within each treatment arm. This is done for five of the models, and one model is simplified to impute across all treatment arms. This latter model was chosen for the ContRAst 1 and 2 studies. The MI models use imputed and observed outcome(s) from preceding visit(s) $Y_{i(j-1)}$ as covariates, with further covariates representing different assumptions about the subset of patients and timepoints that are considered exchangeable with the missing observations. We describe the MI models below using Greek letters for the MI regression coefficients, with $\alpha$ representing intercepts, $\beta$ representing slope parameters for previous values, $\tau$ representing time since discontinuation, $\delta$ or $\gamma$ for indicating patient IE status or pattern.

**Common Intercepts Common Slopes (CICS)**

We use a simple MAR model which does not make a distinction between the pre- and post-IE data. The model to impute the missing outcomes at visit $j$ for each patient $i$ assigned to treatment $z$ *within* treatment arm is:

$$logit[P(Y_{zij} = 1)] = \alpha_{zj} + \beta_{zj0}Y_{zi0} + \cdots + \beta_{zj(j-1)}Y_{zi(j-1)}$$

**On/off Intercepts Common Slopes, pooled by treatment (POOLED OICS)**

This model extends the CICS model by including an indicator for the pre-and post-IE status. In the case of treatment discontinuation, the covariate $D_{ij}$ is included to indicate on- and off-treatment status at the imputation time $j$. Note there are no indexes for $z$ since the missing data are imputed *across* treatment arms.

$$logit[P(Y_{ij} = 1)] = \alpha_j + \boldsymbol{\delta_j D_{ij}} + \beta_{j0}Y_{i0} + \cdots + \beta_{j(j-1)}Y_{i(j-1)}$$

**On/off Intercepts Common Slopes (OICS)**

This model is similar to the POOLED OICS model and imputes the missing data *within* treatment arm. Indices are included for $z$ to capture the effect of assigned treatment varies with the on- and off-treatment status at the imputation time $j$.

$$logit[P(Y_{zij} = 1)] = \alpha_{zj} + \boldsymbol{\delta_{zj} D_{zij}} + \beta_{zj0}Y_{zij0} + \cdots + \beta_{zj(j-1)}Y_{zi(j-1)}$$

**On/off Intercepts and Time Slopes (OITS)**

The OITS model extends the OICS model by including time since discontinuation within assigned treatment. Time since discontinuation is assumed to be linear.

$$logit[P(Y_{zij} = 1)] = \alpha_{zj} + \delta_{zj} \cdot D_{zij} + \boldsymbol{\tau_{zj}} \cdot (J - \boldsymbol{D_{zi}(t)}) + \beta_{zj0}Y_{zi0} + \cdots + \beta_{zj(j-1)}Y_{zi(j-1)}$$

**Pattern Intercepts Common Slopes (PICS)**

The PICS model extends the OICS model by including a set of terms $P_{zi1}, \dots, P_{zij}$ indicating the monotone pattern of treatment discontinuation:

$$logit[P(Y_{zij} = 1)] = \alpha_{zj} + \boldsymbol{\gamma_{zj1}} \cdot \boldsymbol{P_{zi1}} \dots + \boldsymbol{\gamma_{zjj}} \cdot \boldsymbol{P_{zij}} + \beta_{zj0}Y_{zi0} + \cdots + \beta_{zj(j-1)}Y_{zi(j-1)}$$

The CICS, OICS, and PICS models are the binary equivalent to the MI1, MI2 and MI3 models specified by Bell et al(7).

The OICS, POOLED OICS and CICS models were pre-specified for the contRAst 1 and contRAst 2 studies in a step-down approach depending on the availability of post-intercurrent event data (See Example Case Studies in Rheumatoid Arthritis).

## Simulation Set Up

Two simulation studies were conducted to evaluate the performance of the proposed imputation models for estimating the primary estimand in the motivating section. The first simulation study considered different IE rates and missing data within each treatment arm. The second simulation study assessed the performance of the different MI models for smaller vs. larger sample sizes for one scenario from the first simulation study.

## Simulation Study 1: Varying Discontinuation Rates

The first simulation study looked at a fixed sample size of 250 patients per arm. We considered four scenarios that varied the IE rates. Three of these assumed IE rates were equal between treatment arms and one unequal. Under each of these four scenarios, we considered five different withdrawal rates (resulting in missing data). This resulted to a total of 4x5=20 scenarios. For the purpose of assessing any impacts on the Type I error rate, we investigated a null scenario where the on- and off-treatment response rates for the Active arm were set equal to the Control.

## Simulation Study 2: Varying Sample Size

The second simulation study varied sample sizes for 50, 100 and 500 per arm. For this simulation, we explored the scenario where IE rates were unequal between treatment arms, since unequal IE rates are considered more likely to occur in practice. For this scenario, three different withdrawal rates (leading to missing data) were considered. This resulted in 3x1x3=9 additional scenarios. The results from this simulation are compared with the same scenarios set out in the first simulation study for a sample size of 250 patients per arm. As for simulation study 1, we investigated a null scenario where the on- and off-treatment response rates for the Active arm were set equal to the Control to assess the impact on the Type I error rate.

## Data Generation Model (DGM)

Data for patients in each arm were simulated as follows:
1. Standard multivariate normal variables were transformed using the univariate normal CDF to create correlated Gaussian copula values.
2. These values are then transformed to binary indicators of response using the inverse CDF. For each visit (baseline, visit 1, visit 2 and visit 3), binary correlated on-treatment outcomes and (separate) off-treatment outcomes are created. A correlation of 0.5 is assumed across all visits for each treatment arm separately.

Table 1 shows the on-treatment and off-treatment outcome response rates assumed to create the data in Step 1 of the DGM. Since the proportion of patients on-treatment, off-treatment and missing for the contRAst 1 and 2 studies is not publicly available, the on-treatment response rates chosen for the simulation study were approximately based on the proportion of responders(13,17).

ACR20 response is defined as a 20% improvement in symptoms for patients with RA. Improvement cannot be measured at baseline meaning that baseline measures for ACR20 do not exist. Therefore, the proportion of on-treatment response rate at baseline for our simulations was based on the ACR20 response rate seen at Week 1. This was so that all outcome measures generated were assessed on a binary scale.

On-treatment response rates post-baseline for the Control arm were kept at a stable rate of 0.3 rather than creating an on-treatment placebo effect. This was so that any differences subsequently seen in the performance of the models were not driven by the choice of these on-treatment response rates. The off-treatment response rates assume a gradual return to half of the baseline control rate in each arm. This approach was taken to reflect a plausible scenario relating to a placebo effect for RA where patients who were selected to discontinue from treatment had worse outcomes than patients who remain on-treatment on Control. Both simulation studies used the same response rates.

Table 1: Assumed on-treatment and off-treatment response rates.

| Randomised treatment | Baseline | Visit 1 | Visit 2 | Visit 3 |
|---|---|---|---|---|
| **On-treatment response rates** | | | | |
| Active | 0.3 | 0.35 | 0.4 | 0.45 |
| Control | 0.3 | 0.3 | 0.3 | 0.3 |
| **Off-treatment response rates** | | | | |
| Active | | 0.35 | 0.25 | 0.15 |
| Control | | 0.3 | 0.225 | 0.15 |

3. To select patients to discontinue, we use an *intercurrent event at random* (IAR) selection model to indicate patients with an IE at one of the 3 post-baseline visits. To do this, we create a ranked measure ($\kappa_{ij}$) which conditions on the previous on-treatment outcome ($y_{i,j-1}$) and create a monotone discontinued from treatment pattern using:

$$\kappa_{i,j} = \log\left(\frac{v_i}{1-v_i}\right) - \omega_{ij} \cdot y_{i,j-1}$$

where $\omega_{ij} = 0.75$, and $v_i$ are independent draws from $v_i \sim Uniform(0,1)$. The required lowest percentage of patients are selected to discontinue, and their on-treatment outcomes are replaced with their counterfactual simulated off-treatment data. The value of $\omega_{ij}$ was chosen to impose a strong propensity for the previous on-treatment behaviour to drive the discontinuations.

The discontinuation rates in each arm relate to the proportion of patients selected to discontinue treatment by the final timepoint ($Y_3$). The first 3 scenarios are set at equal discontinuation rates for each treatment arm: 10%, 20% and 30%. The fourth scenario looks at unequal rates of discontinuation at 20% on the Control arm and 30% on the Active arm (SI Table 1). The proportion of discontinuations vary over the three post-baseline visits so that more patients are selected to discontinue earlier.

4. As patients cannot withdraw from the study prior to discontinuing treatment, we randomly assign patient withdrawals under a missing completely at random (MCAR) assumption and condition on patients who were selected to discontinue treatment at each visit. This creates an informative missingness mechanism unless the IE in Step 2 is conditioned on the previous on-treatment outcome ($y_{i,j-1}$). To select patients to withdraw from the study under MCAR, we create a measure at each visit for each patient from:

$$u_i \sim Uniform(0,1)$$

These values are then ranked and the patients with the lowest values have their outcomes set to missing only if they experienced an IE at that visit.

The proportion of patients withdrawing from the study post-IE (and are set to be missing) ranges from 30-70% at increments of 10%. For the second simulation study, this ranged between 30-50% to reflect what may be seen in practice.

## MI Analyses and Estimator

A logistic MI model is used to create the imputations for each of the models specified above. Since we are imputing binary data, computational issues may arise as a result of perfect prediction. To avoid issues with perfect prediction in our simulation study, the data are augmented during the imputation procedure for all MI models fitted(18,19). Example SAS code for each of the MI models is included in Appendix 1.

A total of 25 imputations was computationally feasible to create the imputed datasets required for each of the four MI models explored. The results from the analysis model were combined using Rubin's rules(14). Any MI models that could not be fitted were simply excluded from the analysis instead of attempting to collapse them into simpler ones. This was done to assess under which conditions the models may be unsuitable.

Following imputation, the complete (observed and imputed) simulated datasets were analysed using a logistic regression model at the final timepoint including treatment and baseline ($Y_0$) as explanatory variables to estimate the conditional odds ratio. The results from this were compared to results from the full simulated (on- and off-treatment) data, which contains no missing values. As stated in the FDA guideline, it should be prespecified whether the effect of interest in an analysis is conditional or unconditional(20). A conditional odds ratio was chosen to align with the example RA case studies discussed further below. If unconditional (marginal) odds ratios are required for covariate adjustment, methods such as g-computation or propensity scores could be used for analysis(21–23).

All analyses were performed using SAS version 9.4.

## Performance Measures

The performance measures used for each model were bias, 95% confidence interval (CI) halfwidth and coverage, together with power, Type I error and relative error in the model-based standard error (ModSE). Bias is presented as a percentage and calculated in relation to the true log odds. The change in 95% CI halfwidth and coverage was calculated relative to the full simulated data and presented as a percentage change. The proportion of simulated data for which the models converged is also shown. A total of 6000 simulations were performed providing a Monte Carlo Standard Error (MCSE) no larger than 0.001.

The relative error in ModSE is also presented, and defined as the model-based SE relative to the empirical SE (EmpSE), calculated as a percentage. Any decrease in percentage is interpreted as an underestimation of the ModSE.

The Type I error was calculated for a null using a separate set of simulations where the on- and off-treatment response rates for the Active arm were set to be equivalent to the on- and off-treatment response rates for the Control arm.

We note that for the scenario which explores unequal rates of discontinuation (30% Active; 20% Control), the true on-treatment response rates differ to the true off-treatment response rates. Additionally, the proportion of patients off-treatment differ between the active and control arm. This means that the true response rate is also different between the two treatment arms, even under the null. Therefore, for this scenario, the Type I error for the null is not a true Type I error and we do not expect it to equal the nominal 2.5%. For the remainder of this paper we instead refer to this as the "false positive rate".

# Results

## Simulation Study 1: Varying Discontinuation Rates

The CICS, POOLED OICS, OICS and OITS models fitted without any computational issues and therefore all simulations could be fitted for these models. The PICS model ran into issues where ≥60% of patients who had the IE (treatment discontinuation) also withdrew from the study. These issues were more prevalent for the scenario where the discontinuation rate was set to 10% in each arm, since only a small number of patients had observed post-IE data. Table 2 shows the proportion of simulations that could be fitted for the pattern-based PICS model.

Table 2: Number and percentage of simulations fitted for the pattern-based PICS model

| Discontinuation scenario | Withdrawal rate | | | | |
|---|---|---|---|---|---|
| | 30% | 40% | 50% | 60% | 70% |
| 10% Active; 10% Control | 6000 (100%) | 6000 (100%) | 6000 (100%) | 5897 (98.3%) | 4893 (81.6%) |
| 20% Active; 20% Control | 6000 (100%) | 6000 (100%) | 6000 (100%) | 5999 (99.98%) | 5895 (98.25%) |
| 30% Active 30% Control | 6000 (100%) | 6000 (100%) | 6000 (100%) | 6000 (100%) | 5978 (99.6%) |
| 30% Active 20% Control | 6000 (100%) | 6000 (100%) | 6000 (100%) | 5999 (99.98%) | 5948 (99.1%) |

## Bias, Halfwidth and Coverage

Figure 2 presents the bias, change in 95% CI halfwidth and change in CI coverage.

Bias is presented as a percentage on the log odds scale. The FULL model showed a small increase in bias across some scenarios which is due to small sample bias. We confirmed this by running 6000 simulations on 2000 patients per arm for the unequal discontinuation rate scenario (Appendix 2, Figure 1).

The CICS model was consistently biased across all scenarios and the coverage decreased as the rate of discontinuation and amount of missing data increased (Figure 2). The decrease seen in coverage for this model mirrors the scenarios where the bias is much larger. This indicates that the decrease in coverage is driven by the bias. The POOLED OICS model showed very little bias across the majority of scenarios. When discontinuation rates are set equal in both treatment arms, we observe a negligible amount of bias when the missing data are imputed pooled across all treatment arms. This bias increases slightly where there are unequal rates of discontinuation and as the missing data increases. However, even in this scenario, the POOLED OICS model shows far less bias (1-2%) than the CICS model which shows a large amount of bias ranging between 12-32%. The OICS, OITS and PICS models showed no bias or bias that is comparable with the FULL model.

Relative to the FULL model, the change in 95% CI halfwidth showed an increase in width as the complexity of the models increased. This increase is negligible for 30%-50% missing data post-IE but becomes more marked for higher levels of missing data irrespective of the discontinuation rate. Across equal and unequal rates of discontinuation, all models that made a distinction between pre-IE and post-IE data were overcovered for >50% missing data. This is due to the higher amounts of missingness resulting in less observed information to impute from and therefore higher uncertainty of the values being imputed. This is more distinct for the OITS and PICS models which suggests that the inclusion of additional parameters for time and patterns, respectively, further increases the overcoverage.

Figure 2: Simulation results for N= 250 per arm across varying discontinuation rates.

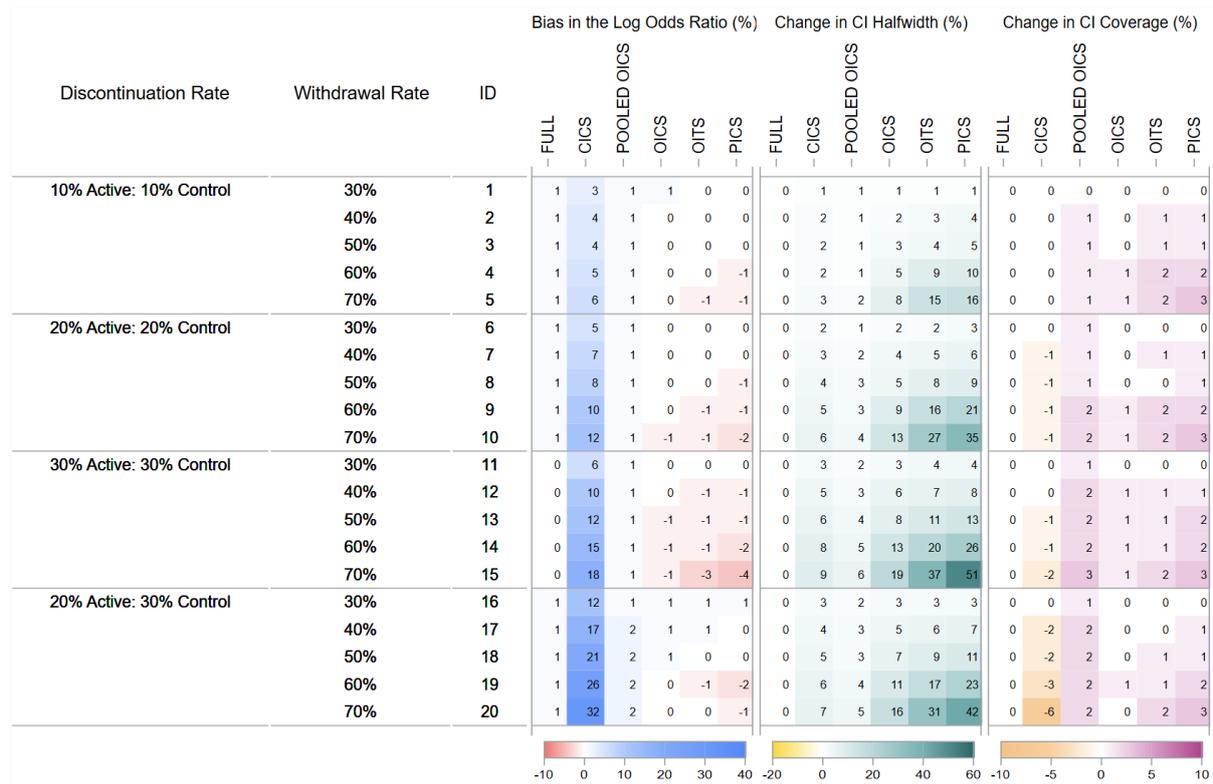

## Power and False Positive Rate

For the FULL model, the power varies depending on the proportion of IEs since the true off-treatment response rates converge to the same value (0.15) on both treatment arms. This means that even when all post-IE data is observed, studies could be underpowered if there is a large proportion of IEs.

Across all discontinuation rates explored, the OICS, OITS and PICS models consistently demonstrated a distinct drop in power compared to the FULL model where less than 50% post-IE data are collected (Figure 3). This is due to the increased uncertainty as the amount of missing data increases. For lower rates of discontinuation and missing data, the power remains above 80%. The CICS model gains power where there are high rates of discontinuation and missingness and this is driven by the increase in bias in the direction of benefit.

Results from looking at the null generally showed that the CICS model has the largest false positive rate compared to the POOLED OICS, OICS, OITS and PICS models. The directional changes here mirror that of the coverage, which suggests that the increase in the false positive rate for the CICS model is due to underestimation of the model-based standard errors relative to the empirical standard error (Table 4). The scenario for the lowest discontinuation rate shows a slight increase in the false positive rate where there are low rates of missing data, and for the FULL model (Table 3). We believe this is due to Monte Carlo error. Increasing the number simulations to 100000 for the FULL model only in this scenario reduced the false positive rate to 2.5% and the relative percentage error in model-based standard errors was 0.42 (results not shown). The POOLED OICS model maintains power that is close to the FULL model, however the false positive rate becomes lower as the rate of discontinuation increases, especially for unequal rates of discontinuation irrespective of the amount of missing data. This is because the model-based standard errors relative to the empirical standard error are overestimated, and this is also reflected in the levels of overcoverage.

Figure 3: Power

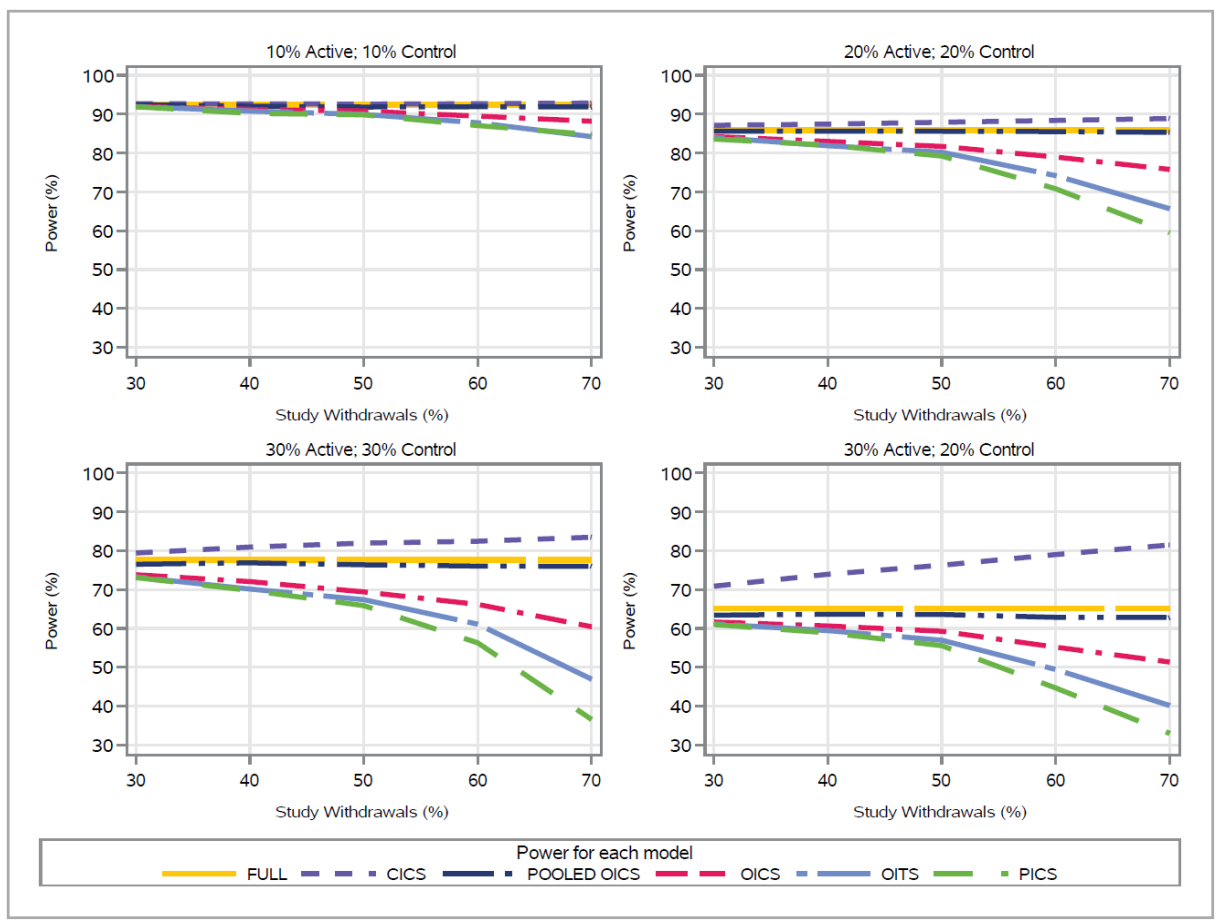

Table 3: False positive rate.

| Scenario | | Models | | | | | |
|---|---|---|---|---|---|---|---|
| Discontinuation rate | Withdrawal rate | FULL | CICS | POOLED OICS | OICS | OITS | PICS |
| **10% Active; 10% Control** | 30 | 2.90 | 2.78 | 2.53 | 2.70 | 2.67 | 2.65 |
| | 40 | 2.90 | 3.00 | 2.65 | 2.67 | 2.67 | 2.55 |
| | 50 | 2.90 | 2.70 | 2.48 | 2.45 | 2.25 | 1.98 |
| | 60 | 2.90 | 2.93 | 2.37 | 2.40 | 1.95 | 1.58 |
| | 70 | 2.90 | 2.80 | 2.30 | 1.90 | 1.48 | 1.18 |
| **20% Active; 20% Control** | 30 | 2.45 | 2.48 | 2.05 | 2.47 | 2.30 | 2.20 |
| | 40 | 2.45 | 2.40 | 1.93 | 2.27 | 2.20 | 2.07 |
| | 50 | 2.45 | 2.53 | 1.87 | 2.63 | 2.37 | 2.10 |
| | 60 | 2.45 | 2.43 | 1.73 | 2.13 | 1.65 | 1.15 |
| | 70 | 2.45 | 2.40 | 1.63 | 1.88 | 1.08 | 0.88 |
| **30% Active; 30% Control** | 30 | 2.50 | 2.57 | 1.95 | 2.43 | 2.10 | 2.07 |
| | 40 | 2.50 | 2.65 | 1.73 | 2.38 | 2.53 | 2.27 |
| | 50 | 2.50 | 2.18 | 1.53 | 2.28 | 2.07 | 1.73 |
| | 60 | 2.50 | 2.45 | 1.47 | 2.43 | 2.15 | 1.52 |

| | 70 | 2.50 | 2.67 | 1.25 | 1.98 | 1.47 | 0.75 |
|---|---|---|---|---|---|---|---|
| **30% Active; 20% Control** | 30 | 1.10 | 1.25 | 0.83 | 1.10 | 1.08 | 1.00 |
| | 40 | 1.10 | 1.40 | 0.72 | 0.98 | 1.00 | 0.95 |
| | 50 | 1.10 | 1.67 | 0.73 | 1.23 | 1.03 | 0.85 |
| | 60 | 1.10 | 1.82 | 0.63 | 1.13 | 0.75 | 0.50 |
| | 70 | 1.10 | 2.03 | 0.52 | 0.93 | 0.65 | 0.34 |

Table 4: Relative % error in model-based standard errors.

| Scenario | | Models | | | | | |
|---|---|---|---|---|---|---|---|
| Discontinuation rate | Withdrawal rate | FULL | CICS | POOLED OICS | OICS | OITS | PICS |
| **10% Active; 10% Control** | 30 | -1.42 | -1.67 | 0.11 | -0.94 | -0.61 | -0.23 |
| | 40 | -1.42 | -2.06 | 1.21 | 0.02 | 1.45 | 2.56 |
| | 50 | -1.42 | -2.40 | 1.10 | -0.18 | 1.80 | 3.64 |
| | 60 | -1.42 | -2.98 | 1.87 | 1.41 | 5.53 | 8.02 |
| | 70 | -1.42 | -3.10 | 2.54 | 4.50 | 11.09 | 13.31 |
| **20% Active; 20% Control** | 30 | 0.67 | -0.54 | 4.05 | 1.05 | 1.38 | 1.72 |
| | 40 | 0.67 | -1.59 | 5.45 | 1.44 | 2.00 | 3.33 |
| | 50 | 0.67 | -2.22 | 6.65 | 1.69 | 3.53 | 5.92 |
| | 60 | 0.67 | -4.13 | 8.00 | 2.72 | 8.11 | 14.64 |
| | 70 | 0.67 | -5.31 | 9.21 | 4.63 | 14.78 | 24.50 |
| **30% Active; 30% Control** | 30 | 1.17 | -0.73 | 5.97 | 1.13 | 1.39 | 1.55 |
| | 40 | 1.17 | -1.88 | 9.11 | 2.40 | 3.05 | 3.97 |
| | 50 | 1.17 | -3.08 | 10.68 | 2.85 | 4.75 | 6.80 |
| | 60 | 1.17 | -5.18 | 12.71 | 3.33 | 6.45 | 13.53 |
| | 70 | 1.17 | -7.90 | 14.54 | 4.93 | 13.65 | 27.79 |
| **30% Active; 20% Control** | 30 | 0.34 | -2.62 | 5.02 | 1.07 | 1.37 | 2.00 |
| | 40 | 0.34 | -5.99 | 6.92 | 1.64 | 2.46 | 3.61 |
| | 50 | 0.34 | -9.32 | 8.06 | 1.87 | 3.31 | 5.71 |
| | 60 | 0.34 | -12.25 | 10.32 | 3.55 | 8.29 | 14.55 |
| | 70 | 0.34 | -17.45 | 11.01 | 4.16 | 13.40 | 25.23 |

## Simulation Study 2: Varying Sample Size

Under a range of different sample sizes, the CICS, POOLED OICS, OICS and OITS models ran without any computational issues. For a sample size of 100 patients per arm, when there was 50% missing post-IE data the PICS model could not be fitted for 10 (0.2%) simulations (SI Table 2).

### Bias, Halfwidth and Coverage

Figure 4 presents the bias, change in 95% CI halfwidth and change in CI coverage. The change in 95% CI halfwidth and coverage was calculated relative to the full simulated data (FULL results).

We note that the FULL model was slightly biased. This bias diminishes as the sample size increases and provides reassurance that the bias is driven by the small sample size which is common for binary data[24–26]. Similar to the results when looking at 250 patients per arm, the CICS model was the most biased irrespective of sample size (Figure 4). For smaller sample sizes (N=50 per arm), some bias was shown in the POOLED OICS, OICS, OITS and PICS models, however these models are comparable and well-aligned to the FULL model.

Consistent with the results shown previously for 250 patients per arm, the change in CI halfwidth increases as the complexity of the model increases relative to the FULL across the different sample sizes, especially when there is 40-50% missing data post-IE.

For larger sample sizes (N≥250 per arm), coverage was well-maintained for the OICS and OITS models and comparable to the FULL model. Coverage for the PICS model increased slightly as the amount of missing data increased across all sample sizes. For the smaller sample sizes (N≤100 per arm), the POOLED OICS, OICS, OITS and PICS models were slightly overcovered and the CICS model appeared to show adequate coverage. Bias of the CICS model consistently dominates and increases as the proportion of missing data increases, it is likely that the model-based standard errors are underestimated compared with the empirical standard errors (SI Table 3). This is reflected by the fact that the CICS model becomes undercovered, especially when there is 40-50% missing data for a sample size of 500 patients per arm.

Figure 4: Simulation results for N= 250 per arm across varying discontinuation rates.

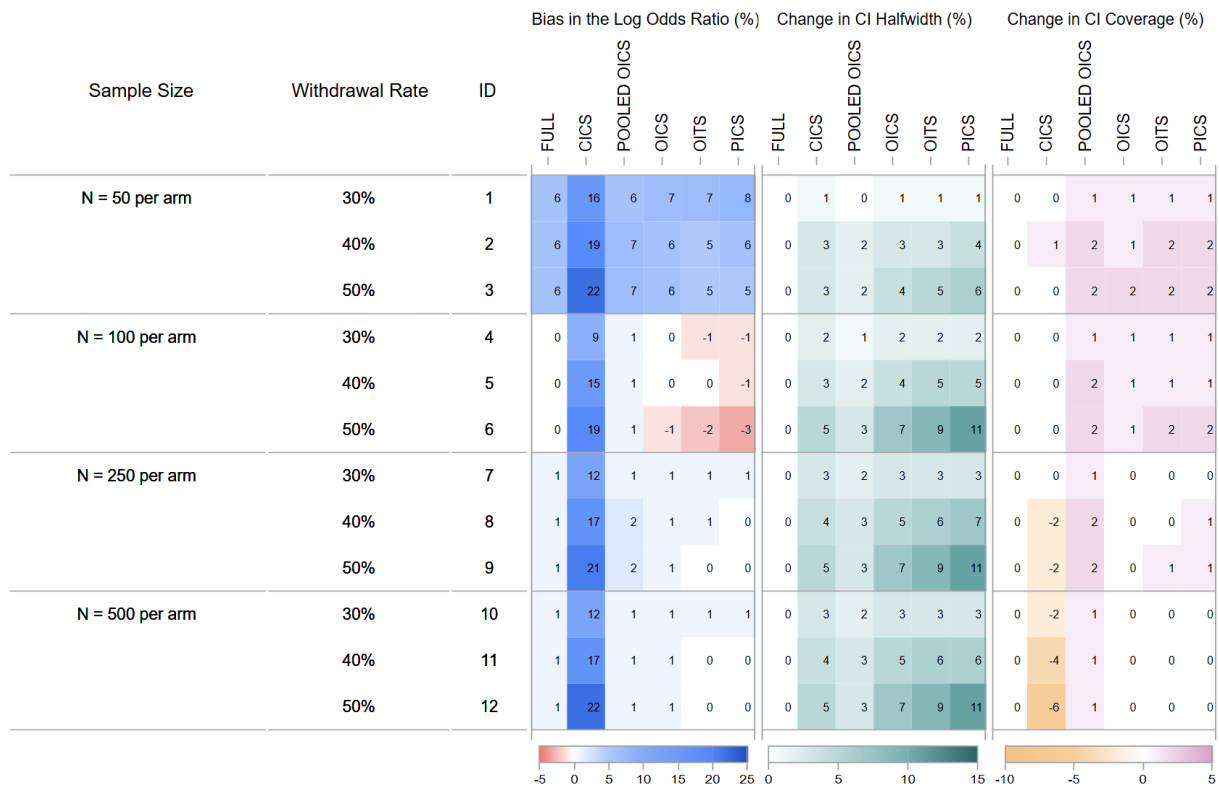

## Power and False Positive Rate

As the number of patients per arm increases, the power increases (SI Figure 1). Across all sample sizes explored, the power for the POOLED OICS, OICS, OITS and PICS models were similar. For large sample sizes (N=500 per arm), the drop in power for the OICS, OITS and PICS models is negligible when compared to the FULL model even up to 50% missing data post-IE. Consistent with the first simulation study where there are 250 patients per arm, the power for the CICS model was higher than the FULL model which is a result of the increased bias. The false positive rate is consistent with what was seen in Simulation Study 1 and remains low for all sample sizes explored (SI Table 4). The CICS model showed that the false positive rate was consistently higher when compared to the FULL model and the POOLED

OICS model was consistently lower when compared to the FULL model. This is reflected within the CI coverage and model-based standard errors relative to the empirical standard error (SI Table 3).

## Further investigations of the OICS model

The results from the above two simulation studies show the OICS model could be a promising approach to impute missing data post-IE. To further stress-test the OICS model, we performed an additional simulation study that varied the on- and off-treatment response rates (SI: Further investigations of the OICS model). We looked at a very low, a medium and a high response rate, and compared the results from this to the original response rates in Simulation Study 1 to see the impact on performance measures. We found that the OICS model performs poorly in cases where there are large differences between the on- and off-treatment response rates. For further detail see the SI (SI Figures 2-5).

## Example Case Studies in Rheumatoid Arthritis

For the motivating examples contRAst 1 and contRAst 2, a pre-specified step-down approach to impute the missing off-treatment data was defined. Missing data for the ACR20 response resulting from study withdrawal were imputed using one of three possible MI models depending on the availability of off-treatment (post-IE) data which is described in the publication's Supplementary Appendix(13). We provide a brief summary below:

**Option 1: MI using off-treatment data within randomised treatment arm:**
If there was sufficient off-treatment data within a randomised treatment, then missing data would be imputed conditional on the participant's observed outcomes, using off-treatment data within randomised treatment arm. This option is equivalent to the OICS model defined in the simulation studies.

**Option 2: MI using off-treatment data across all treatment arms:**
If the model in option 1 could not be fitted due to insufficient availability of the off-treatment data, then the data across all treatment arms would be combined. This option uses an MI model that is similar the OICS model we defined for the simulation study but differs in that the treatment arms were pooled to impute the missing data rather than imputed by treatment arm. This option is equivalent to the POOLED OICS model.

**Option 3: MI under MAR assumption:**
If option 1 and 2 could not be fitted due to insufficient off-treatment data, then all missing data would be imputed under a MAR assumption. This is equivalent to the CICS model defined above.

Table 5: Odds ratio for ACR20 from contRAst 1 and conRAast 2.

| n (%) | Pooled Placebo N=256 | OTI 90 mg QW (N=513) | OTI 150 mg QW (N=510) |
|---|---|---|---|
| **contRAst 1** | | | |
| Responders, % | 42.7 | 54.7 | 50.9 |
| Odds ratio (95%) CI | | 1.62 (1.19 to 2.21) | 1.39 (1.02 to 1.89) |
| **contRAst 2** | | | |
| Responders, % | 32.5 | 54.9 | 54.5 |
| Odds ratio (95%) CI | | 2.57 (1.87 to 3.53) | 5.38 (3.66 to 7.90) |

Both the contRAst 1 and contRAst 2 studies used the second option to impute missing data since there was insufficient post-IE information to impute within the randomised treatment arm, leading to non-convergence of the MI model (option 1). Table 5 shows the results from each study following the successful implementation of using MI to account for the occurrence of the IE(13).

## Discussion

We have proposed and investigated different models using multiple imputation for binary outcomes under a range of scenarios to estimate the effect of interest where a single intercurrent event is handled using a treatment policy strategy. In addition, we have highlighted how these models can be implemented in practice illustrated by the contRAst 1 and 2 studies. To our knowledge, this is the first time these types of models have been explored for a binary endpoint to address missingness when partial post-IE data exists.

The simulation study demonstrated that using a basic MAR model (CICS) that makes no distinction between the pre-IE and post-IE data can be substantially biased and have poor coverage. Conversely, the POOLED OICS, OICS, OITS and PICS models, which account for the occurrence of the IE, show negligible bias, and have fairly well-maintained CI coverage up to 50% missing data post-IE. When there is more than 50% missing data post-IE, the POOLED OICS, OICS, OITS and PICS models showed overcoverage. This illustrates that using an estimation method that relies on a basic MAR assumption is a poor choice when targeting an estimand using a treatment policy strategy. A distinction needs to be made between the pre- and post-IE data and this should be reflected appropriately in MI models to impute the missing data.

The POOLED OICS model showed small levels of bias when IE rates are equivalent between treatment arms and is more biased with slight overcoverage when the IE rates are unequal between treatment arms. Even in this case, including an indicator for the IE and combining treatment arms when retrieved dropout MI models cannot be fitted within assigned treatment arms still appears to be a better choice than a basic MAR model. However, this may come at a cost of using an overly conservative model.

It is important to note that even when there is no missing data, power decreases as the proportion of IEs increases. As the missing data also increases, the power drops for all MI models explored which account for the post-IE information within each treatment arm. This highlights the importance to consider the expected rates of IEs and amount of post-IE information that can be collected since the choice between using the OICS, OITS and PICS model may not be a trivial one. When it is not possible to determine expected rates of IEs and missing data, adopting a step-down approach depending on the availability of post-IE data could be used. This was the approach taken for the contRAst studies, where the most complex but realistic model was planned for, on the condition there were enough patients with available post-IE data to impute from.

We also looked at different sample sizes for unequal rates of discontinuation for 30-50% missing data post-IE, which we considered to be the most realistic scenario the majority of study teams will encounter. Irrespective of the sample size, the findings from this simulation study are well-aligned with the results from the first simulation study.

Regardless of sample size, the OICS model seems to perform adequately if having a small amount of bias for unequal rates of discontinuation (the most realistic scenario) is deemed acceptable up to 50% missing post-IE data, and provided that the off-treatment response rates are not too extreme from the on-treatment response rates. For studies with a long duration, where the effect of treatment gradually improves compared to the control, and after the occurrence of an IE, the effect in the active arm is expected to slowly align with the trajectory observed in the control arm, the OICS model is likely to be biased. This is because the OICS model averages across all IE patterns when it is important to capture these effects over time. In these cases, the PICS model may be a better option.

With a larger sample size, the choice between an OICS, OITS or PICS model seems to matter less up to 40-50% missing data post-IE. In these instances, using the OICS model is the preferred choice due to the simplicity and computational speed of fitting the model. However, if there are large amounts of missing data post-IE and a large enough sample size, the PICS model could be a more suitable choice for estimation provided there are a smaller number of visits to impute from.

The PICS model runs into complications where there are large amounts of missing data. This is because if there is sparse observed data available within the different patterns, the model cannot be fitted. As the number of visits increases, there is a higher chance that the patterns will become inestimable due to small amounts of post-IE data. This is an additional aspect to consider when using this model. A practical solution for this is to collapse the patterns to enable fitting as shown by Bell et al.(7), however, depending on how far the collapsibility needs to go, could lead to using a basic MAR model in some situations.

It is important to note that where there were computational issues with fitting the PICS model, the simulation was excluded. Although the PICS model showed a low false positive rate, indicating no issues from using this model, this needs to be carefully interpreted since the calculation only included the simulations where the model could be fitted. A more thorough look at the false positive rate across these models would include pattern merging. However, where there were no fitting issues found for the PICS model, it is reassuring that the false positive rate remained consistently low irrespective of sample size for the unequal discontinuation rate scenario when there is 30-50% missing data post-IE.

The models explored here included imputing the missing outcomes for 3 post-baseline visits. For a larger number of post-baseline visits, the issue of perfect prediction may still be an issue even after augmentation(27). Further exploration of this to resolve the issues around perfect prediction would be interesting.

At least 50% of post-IE data needs to be recovered for the OICS, OITS and PICS models explored here to produce sensible results. On further investigation (Appendix 3), consistent with results found by Drury et al. and Bell et al. for continuous endpoints(6,7) and outlined in Roger et al.(3), the relative increase in the variance of the estimator due to missing data increases as the proportion of patients with missing off-treatment data increases. We echo the point made by those authors that during study conduct, the efforts made by study teams to continue collecting post-IE data are vital to ensure accurate estimation when these types of models are used to estimate estimands when using a treatment policy strategy to handle IEs. Additionally, considering rates of the IEs and amount of post-IE data that can be collected is important when considering the sample size for trials, since obtaining a low amount of post-IE information where lots of IEs are expected could inadvertently lead to an underpowered study.

The simulation study explored here was performed for binary outcomes. The results from Bell et al.(7) suggest that reference-based options might be more viable to estimate the estimand where IEs are handled using a treatment policy strategy provided the reference-based assumptions for the missing data made are sensible. The work recently published by Cro et al.(28) uses retrieved dropout reference-based centred MI. The methodology uses a Bayesian framework to extend reference-based multiple imputation for continuous data. Using this type of model structure would be interesting for binary outcomes. The proposed models also rely on having adequate post-IE information to impute from and a sensible choice of variance for the informative prior. With no post-IE data, the models revert to a standard reference-based model and a poor, misinformed choice about the variance for the informative prior could alter results. The methodology here is promising, and further research using Bayesian methods could be interesting for other endpoint types.

## Conclusion

Consistent with other research, estimation of estimands where IEs are handled using a treatment policy strategy is challenging. The models proposed here can be used for estimation under certain conditions and accepting a small amount of bias, which is arguably a better option compared with having a large amount of bias when using a basic MAR model that is often assumed.

The choice of model will likely depend on the disease area and expected off-treatment response rates in each arm. We reiterate the importance of study conduct to ensure data continues to be collected post-IE when a treatment policy strategy is specified to handle IEs.

More complex pattern models may be better for larger sample sizes and are worth considering. From the models we explored here, we recommend the OICS model as the most pragmatic option as it is the simplest model which accounts for post-IE data and is unlikely to present fitting problems for the majority of studies. In situations where this model cannot be fitted due to sparse post-IE data or computational issues, we recommend imputing the missing data across all treatment arms and including an indicator for the IE status along with reporting the proportion of IEs by treatment arm.

# Appendix 1: Example SAS code

**POOLED OICS**
```
proc mi data    = ds
        out     = mi_m_y
        nimpute = 25
        seed    = 12345;
  by    all tdratecd wrate scenariocd sim;
  class d1-d3 y0-y3;
  var   d1-d3 y0-y3 ;
  monotone logistic (y1 = d1 y0 / likelihood = augment);
  monotone logistic (y2 = d2 y0 y1 / likelihood = augment);
  monotone logistic (y3 = d3 y0 y1 y2 / likelihood = augment);
run;
```

**OICS**
```
proc mi data    = ds
        out     = mi_m_y
        nimpute = 25
        seed    = 12345;
  by    all tdratecd wrate scenariocd sim trt;
  class d1-d3 y0-y3;
  var   d1-d3 y0-y3 ;
  monotone logistic (y1 = d1 y0 / likelihood = augment);
  monotone logistic (y2 = d2 y0 y1 / likelihood = augment);
  monotone logistic (y3 = d3 y0 y1 y2 / likelihood = augment);
run;
```

**OITS**
```
proc mi data    = ds
        out     = mi_m_y (rename = (_imputation_ = imputation))
        nimpute = &nimp.
        seed    = 12345;
  by    all tdratecd wrate scenariocd sim trt;
  class d1-d3 y0-y3;
  var   timed d1-d3 y0-y3;
  monotone logistic (y1 = timed d1 y0 / likelihood = augment);
  monotone logistic (y2 = timed d2 y0 y1 / likelihood = augment);
  monotone logistic (y3 = timed d3 y0 y1 y2 / likelihood = augment);
run;
```

**PICS**
```
proc mi data    = ds  (where = (no_mi_flag ne 1))
        out     = mi_m_y (rename = (_imputation_ = imputation))
        nimpute = &nimp.
        seed    = 12345;
  by    all tdratecd wrate scenariocd sim trt;
  class p1-p3 y0-y3;
  var   p1-p3 y0-y3;
  monotone logistic (y1 = p1 y0 / likelihood = augment);
  monotone logistic (y2 = p2 y0 y1 / likelihood = augment);
  monotone logistic (y3 = p3 y0 y1 y2 / likelihood = augment);
run;
```

# Appendix 2: Small sample bias

To investigate whether some of the bias seen in the FULL model in Figure 2 where we have unequal discontinuation rates (30% Active; 20% Control) is due to a small sample size, we performed a smaller simulation for 2000 patients per arm and checked the bias up to 50% missing data. A total of 6000 simulations were performed.

Appendix 2, Figure 1: Bias for N=2000 per arm

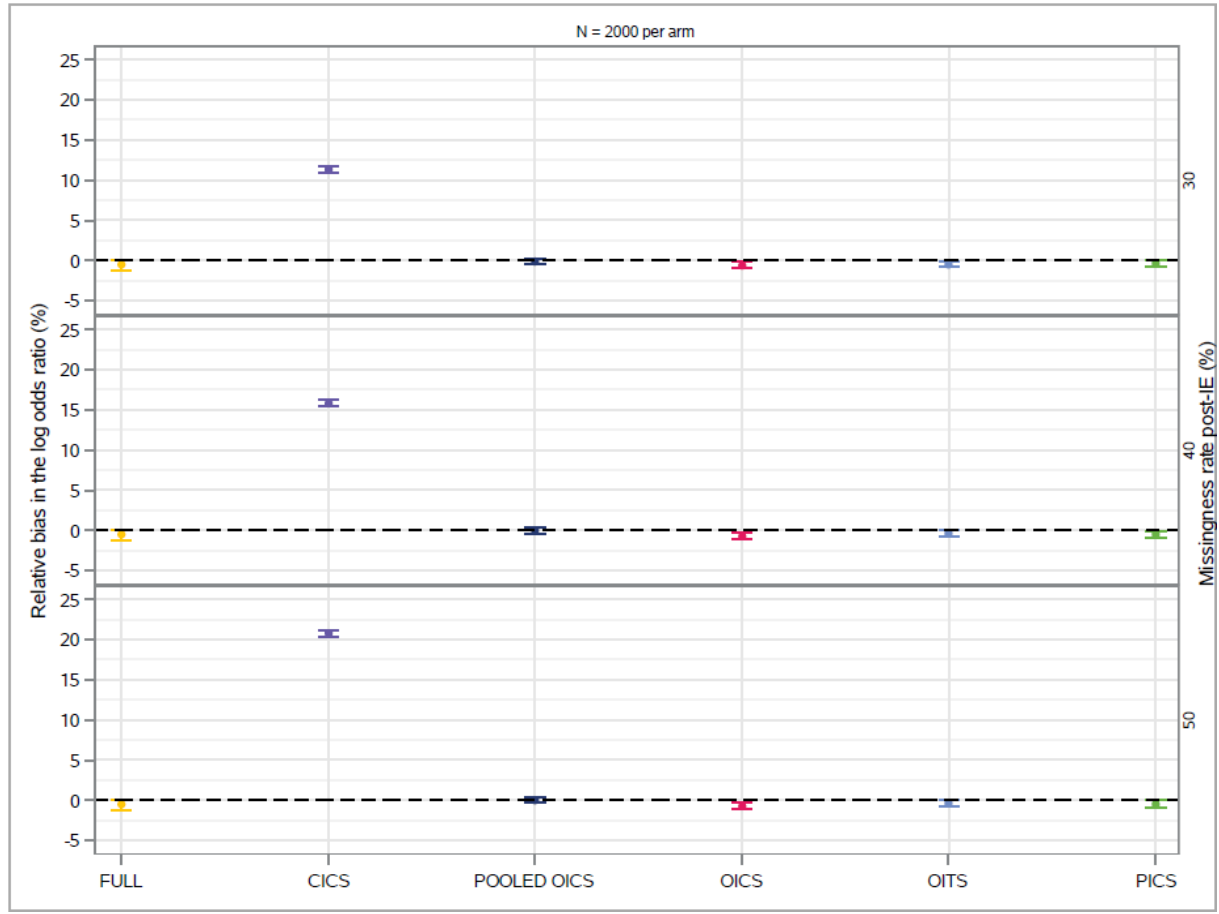

# Appendix 3: Variance inflation

We consider a simple setting for a single treatment arm and the proportion of success follows a Bernoulli distribution. Patients can be categorised in three groups:

The first group are of patients that complete the study and remain on-treatment throughout, denoted by $n_1$. The proportion of patients observed at the timepoint assessment is $p_1$ with variance $p_1(1-p_1)$.

The second group of patients are those that complete the study but discontinue at some point over the duration of the study therefore providing outcome measures while on- and off-treatment. The proportion of patients observed in this group at the primary timepoint assessment are $p_2$ with variance $p_2(1-p_2)$ and number of patients in this groups are denoted by $n_2$.

The third group of patients are those that discontinue over the course of the study and subsequently withdraw, failing to complete the study. This group of patients have information on outcome measures while on-treatment and *potentially* some off-treatment information prior to withdrawal (resulting in subsequent missing data). The number of patients in this groups are denoted by $n_3$ and proportion of patients in this group at the primary timepoint assessment are *not observed*.

The total number of patients is: $N = n_1 + n_2 + n_3$, and assuming those in group 2 are similar to those in group 3 (i.e. $p_3 = p_2$) the trial proportions are:

$$\pi_1 = \frac{n_1 p_1}{n_1 + n_2 + n_3}, \pi_2 = \frac{n_2 p_2}{n_1 + n_2 + n_3}, \pi_3 = \frac{n_3 p_2}{n_1 + n_2 + n_3}$$

And the estimated proportion at the timepoint assessment would be:

$$\mathbb{E}[Y_{policy}] = \pi_{policy} = \frac{n_1 p_1 + (n_2 + n_3) p_2}{n_1 + n_2 + n_3}$$

Assuming independence between the proportions in group 1 and group 2 & 3, and also assuming that $Var(\pi_2) = Var(\pi_3)$ the variance would be:

$$\frac{n_1^2}{(n_1 + n_2 + n_3)^2} Var(\pi_1) + \frac{(n_2 + n_3)^2}{(n_1 + n_2 + n_3)^2} Var(\pi_2)$$

The variance for those that complete the study without discontinuing is:

$$var(\pi_1) = \frac{p_1(1 - p_1)}{n_1}$$

When all data are observed and there is no missing data, the variance is:

$$var(\pi_2) = \frac{p_2(1 - p_2)}{n_2 + n_3}$$

And in the presence of missing data there is no information for patients in group $n_3$, therefore the variance is simply:

$$var(\pi_2) = \frac{p_2(1 - p_2)}{n_2}$$

When <u>all</u> data are observed, the variance of the estimator is:

$$\frac{n_1^2}{(n_1 + n_2 + n_3)^2} \cdot \frac{p_1(1 - p_1)}{n_1} + \frac{(n_2 + n_3)^2}{(n_1 + n_2 + n_3)^2} \cdot \frac{p_2(1 - p_2)}{n_2 + n_3}$$

$$= \frac{n_1 p_1(1 - p_1)}{(n_1 + n_2 + n_3)^2} + \frac{(n_2 + n_3) \cdot p_2(1 - p_2)}{(n_1 + n_2 + n_3)^2}$$

$$= \frac{n_1 p_1(1 - p_1) + (n_2 + n_3) \cdot p_2(1 - p_2)}{n^2}$$

$$= \frac{1}{n^2}[n_1 p_1(1 - p_1) + (n_2 + n_3) \cdot p_2(1 - p_2)]$$

When there is <u>missing</u> data, the variance of the estimator is:

$$\frac{n_1^2}{(n_1 + n_2 + n_3)^2} \cdot \frac{p_1(1 - p_1)}{n_1} + \frac{(n_2 + n_3)^2}{(n_1 + n_2 + n_3)^2} \cdot \frac{p_2(1 - p_2)}{n_2}$$

$$= \frac{n_1 p_1(1 - p_1)}{(n_1 + n_2 + n_3)^2} + \frac{(n_2 + n_3)^2 \cdot p_2(1 - p_2)}{n_2(n_1 + n_2 + n_3)^2}$$

$$= \frac{1}{n^2}\left(n_1 p_1(1 - p_1) + \frac{(n_2 + n_3)^2 \cdot p_2(1 - p_2)}{n_2}\right)$$

$$= \frac{1}{n^2}\left[n_1 p_1(1 - p_1) + p_2(1 - p_2)\left(n_2 + 2n_3 + \frac{n_3^2}{n_2}\right)\right]$$

The additional variance can be worked out as:

$$\frac{1}{n^2}\left[n_1 p_1(1 - p_1) + p_2(1 - p_2)\left(n_2 + 2n_3 + \frac{n_3^2}{n_2}\right)\right] - \frac{1}{n^2}[n_1 p_1(1 - p_1) + (n_2 + n_3) \cdot p_2(1 - p_2)]$$

$$= \frac{p_2(1 - p_2)}{n^2}\left[n_2 + 2n_3 + \frac{n_3^2}{n_2} - (n_2 + n_3)\right]$$

$$= \frac{p_2(1 - p_2)}{n^2} \cdot \left(\frac{n_3^2}{n_2} + n_3\right)$$

$$= \frac{n_3 p_2(1 - p_2)}{n^2} \cdot \left(1 + \frac{n_3}{n_2}\right)$$

$$= p_2(1 - p_2) \cdot \left(1 + \frac{n_3}{n_2}\right) \cdot \frac{n_3}{n^2}$$

The relative increase in variance is:

$$\frac{p_2(1 - p_2) \cdot \left(1 + \frac{n_3}{n_2}\right) \cdot \frac{n_3}{n^2}}{\frac{n_1 p_1(1 - p_1) + p_2(1 - p_2)(n_2 + n_3)}{n^2}} = \frac{n_3 p_2(1 - p_2)}{\cancel{n^2}}\left(1 + \frac{n_3}{n_2}\right) \cdot \frac{\cancel{n^2}}{n_1 p_1(1 - p_1) + p_2(1 - p_2)(n_2 + n_3)}$$

$$= \frac{n_3 p_2(1 - p_2)\left(1 + \frac{n_3}{n_2}\right)}{n_1 p_1(1 - p_1) + p_2(1 - p_2)(n_2 + n_3)}$$

If we assumed that $p_1 = p_2 = p$ and, therefore, the variance $p_1(1 - p_1) = p_2(1 - p_2) = p(1 - p)$. Then:

$$\frac{n_3 p(1 - p)\left(1 + \frac{n_3}{n_2}\right)}{p(1 - p)(n_1 + n_2 + n_3)} = \frac{n_3}{n} \cdot \left(1 + \frac{n_3}{n_2}\right)$$

Which is equivalent to the variance inflation for continuous endpoints.

Consistent with what was observed in the continuous, recurrent event and time-to-event settings, the increase in variance is a consequence of missing data. The relative increase in the variance of the estimator due to missing data increases as the proportion of those with missing data increases ($n_3$) but exacerbated when $n_2$ is small relative to the number missing $n_3$.

# Supplementary Information

SI Table 1: Scenarios for discontinuation rates

| Discontinuation rates | Treatment arm | Visit 1 | Visit 2 | Visit 3 |
|---|---|---|---|---|
| 10% Active; 10% Control | Active | 5% | 3% | 2% |
| | Control | | | |
| 20% Active; 20% Control | Active | 10% | 6% | 4% |
| | Control | | | |
| 30% Active; 30% Control | Active | 15% | 9% | 6% |
| | Control | | | |
| 30% Active; 20% Control | Active | 15% | 9% | 6% |
| | Control | 10% | 6% | 4% |

SI Table 2: Proportion of simulations fitted for the pattern model

| N per arm | Withdrawal rate | | |
|---|---|---|---|
| | 30% | 40% | 50% |
| 50 | 6000 (100%) | 6000 (100%) | 6000 (100%) |
| 100 | 6000 (100%) | 6000 (100%) | 5990 (99.8%) |
| 250 | 6000 (100%) | 6000 (100%) | 6000 (100%) |
| 500 | 6000 (100%) | 6000 (100%) | 6000 (100%) |

SI Table 3: Relative % error in model-based standard errors for varying sample sizes.

| N per arm | Withdrawal rate | FULL | CICS | POOLED OICS | OICS | OITS | PICS |
|---|---|---|---|---|---|---|---|
| 50 | 30 | -1.35 | -2.44 | 0.940 | -0.58 | -0.08 | 0.11 |
| | 40 | -1.35 | -2.02 | 3.989 | 1.83 | 3.28 | 4.23 |
| | 50 | -1.35 | -2.32 | 5.329 | 3.00 | 5.03 | 6.72 |
| 100 | 30 | 1.65 | 0.74 | 5.065 | 2.42 | 3.09 | 3.76 |
| | 40 | 1.65 | -0.70 | 7.175 | 3.59 | 4.82 | 6.07 |
| | 50 | 1.65 | -1.68 | 8.528 | 4.46 | 7.67 | 9.84 |
| 250 | 30 | 0.34 | -2.62 | 5.022 | 1.07 | 1.37 | 2.00 |
| | 40 | 0.34 | -5.99 | 6.916 | 1.64 | 2.46 | 3.61 |
| | 50 | 0.34 | -9.32 | 8.056 | 1.87 | 3.31 | 5.71 |
| 500 | 30 | 2.85 | -4.78 | 6.471 | 2.68 | 2.74 | 2.94 |
| | 40 | 2.85 | -10.59 | 7.884 | 2.83 | 3.12 | 3.82 |
| | 50 | 2.85 | -15.71 | 9.491 | 3.16 | 3.99 | 5.63 |

SI Figure 1: Power for varying sample sizes

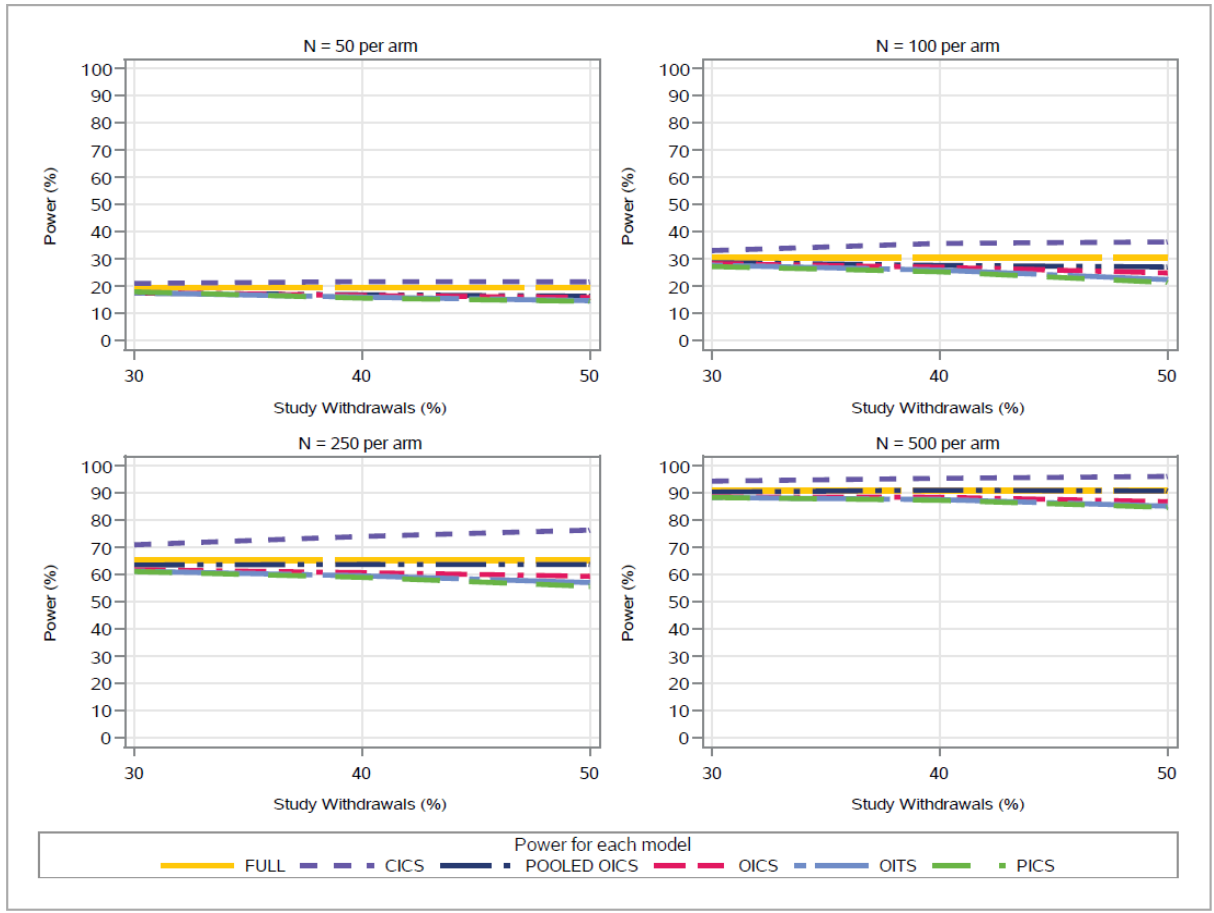

SI Table 4: False positive rate

| N per arm | Withdrawal rate | FULL | CICS | OICS POOLED | OICS | OITS | PICS |
|---|---|---|---|---|---|---|---|
| 50 | 30 | 1.95 | 2.10 | 1.65 | 1.95 | 1.82 | 1.85 |
| | 40 | 1.95 | 2.08 | 1.43 | 1.80 | 1.63 | 1.50 |
| | 50 | 1.95 | 1.93 | 1.33 | 1.53 | 1.40 | 1.20 |
| 100 | 30 | 1.22 | 1.22 | 1.00 | 1.17 | 1.17 | 1.18 |
| | 40 | 1.22 | 1.40 | 0.85 | 1.03 | 1.05 | 0.92 |
| | 50 | 1.22 | 1.35 | 0.77 | 1.10 | 1.07 | 0.83 |
| 250 | 30 | 1.10 | 1.25 | 0.83 | 1.10 | 1.08 | 1.00 |
| | 40 | 1.10 | 1.40 | 0.72 | 0.98 | 1.00 | 0.95 |
| | 50 | 1.10 | 1.67 | 0.73 | 1.23 | 1.03 | 0.85 |
| 500 | 30 | 0.58 | 0.80 | 0.33 | 0.50 | 0.52 | 0.50 |
| | 40 | 0.58 | 0.88 | 0.37 | 0.62 | 0.70 | 0.57 |
| | 50 | 0.58 | 0.97 | 0.38 | 0.62 | 0.58 | 0.53 |

# SI: Further investigations of the OICS model

So far, the OICS model appears a promising choice to impute post-IE missing data irrespective of sample size and up to 50% missing data. However this is under a specific set of assumed response rates. We look at a range of different on-treatment and off-treatment response rates to stress-test under what conditions the OICS model may be unsuitable for.

To do this, we performed a further set of simulations for the unequal rates of discontinuation scenario (30% Active and 20% Control). Only this scenario was chosen as unequal rates of discontinuations are most likely to typically occur in clinical trials. Different rates of missing data ranged between 30%-70%. A total of 6000 simulations were performed.

The various on-treatment and off-treatment response rates are displayed below (SI Table 5).

SI Table 5: Assumed on-treatment and off-treatment response rates to stress-test the OICS model.

|  | **On-treatment response rates** | | | | **Off-treatment response rates** | | |
| --- | --- | --- | --- | --- | --- | --- | --- |
|  | Baseline | Visit 1 | Visit 2 | Visit 3 | Visit 1 | Visit 2 | Visit 3 |
| **High response rate** | | | | | | | |
| Active | 0.8 | 0.85 | 0.875 | 0.9 | 0.8 | 0.6 | 0.4 |
| Control | 0.8 | 0.8 | 0.8 | 0.8 | 0.7 | 0.55 | 0.4 |
| **Medium response rate** | | | | | | | |
| Active | 0.5 | 0.55 | 0.6 | 0.7 | 0.5 | 0.4 | 0.25 |
| Control | 0.5 | 0.5 | 0.5 | 0.5 | 0.45 | 0.375 | 0.25 |
| **Low response rate** | | | | | | | |
| Active | 0.1 | 0.125 | 0.15 | 0.2 | 0.1 | 0.08 | 0.05 |
| Control | 0.1 | 0.1 | 0.1 | 0.1 | 0.09 | 0.075 | 0.05 |

The OICS model performs poorly and becomes biased downwards for the scenario that looked at high response rates (SI Figure 2). Investigating this further, we compared the "High response rate" results to a situation where the off-treatment rate was set to be the baseline control rate (0.8) by the final visit (SI Figure 3). In this scenario the bias is smaller and the variance in the point estimates is also smaller. This suggests that the poor performance of the OICS model in the "High response rate" scenario is driven by large differences between the on- and off-treatment response rates, rather than the high on-treatment response rates.

SI Figure 2: Relative bias on log odds scale for different on- and off-treatment response rate scenarios.

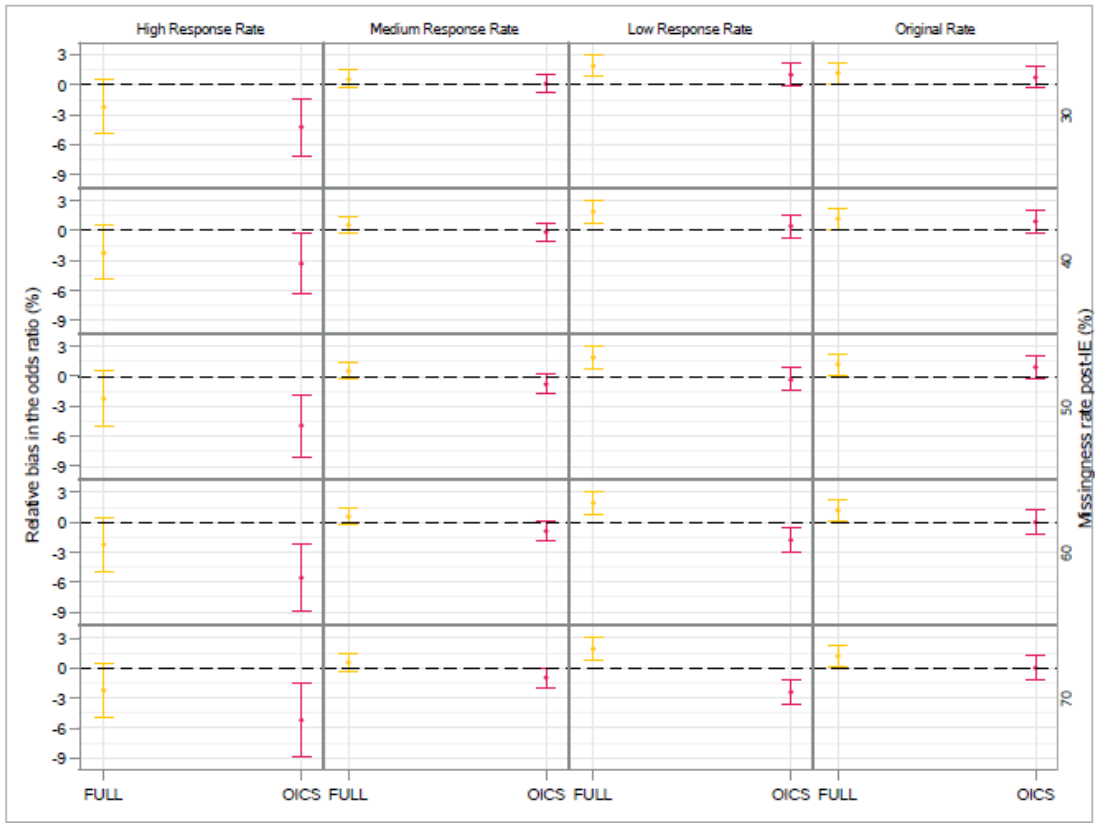

SI Figure 3: Relative bias on the log odds scale for the high response rate scenario.

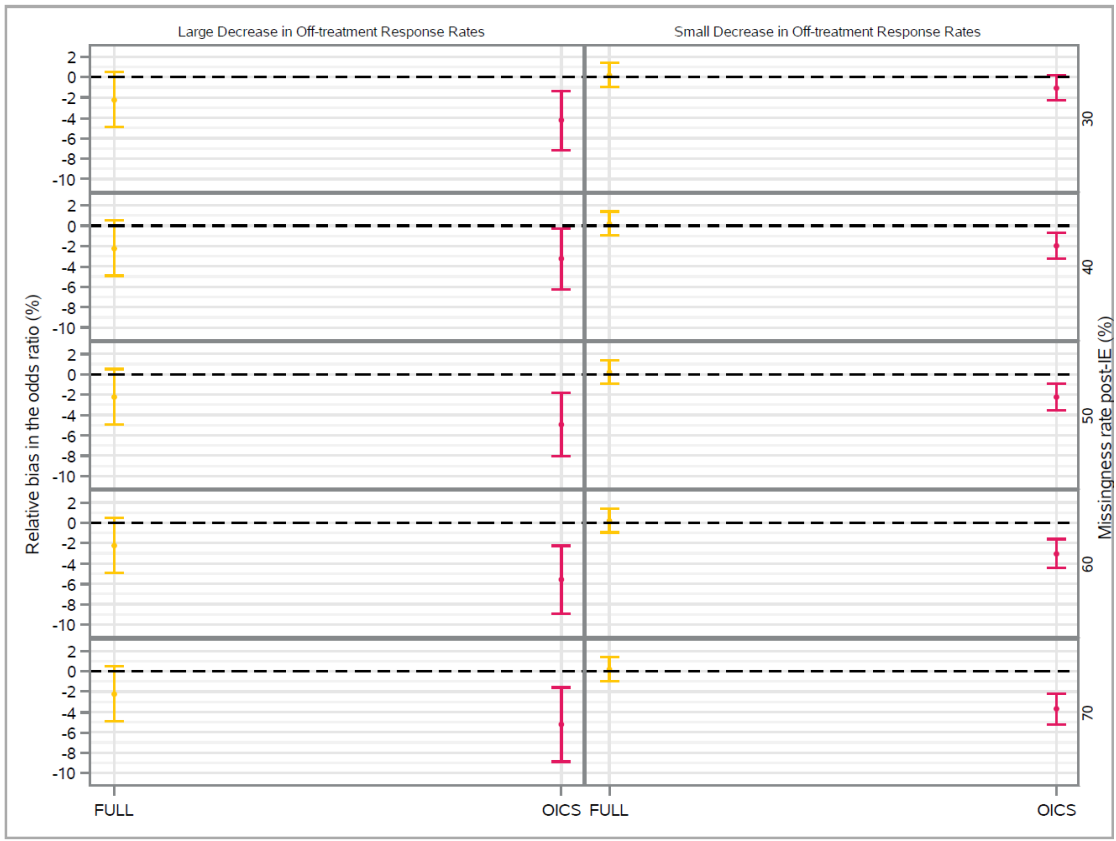